\documentclass[prl,twocolumn,amsmath,amssymb,letterpaper,superscriptaddress,showpacs]{revtex4}

\usepackage{bm}
\usepackage{mathptmx} 

\usepackage{graphicx}

\begin{document}

\title{Experimental confirmation of tissue liquidity based on the exact solution
  of the Laplace equation}

\author{Cyrille Norotte} 
\affiliation{Department of Biological Sciences, University of Missouri, Columbia, MO 65211}

\author{Francoise Marga} 
\affiliation{Department of Physics \& Astronomy, University of Missouri, Columbia,
  MO 65211}

\author{Adrian Neagu} 
\affiliation{Department of Physics \& Astronomy, University of Missouri, Columbia,
  MO 65211}
\affiliation{Department of Biophysics and Medical Informatics, University of
  Medicine and Pharmacy Timisoara, 300041 Timisoara, Romania}

\author{Ioan Kosztin}
\email{kosztini@missouri.edu} 
\affiliation{Department of Physics \& Astronomy, University of Missouri, Columbia,
  MO 65211}

\author{Gabor Forgacs}
\email{forgacsg@missouri.edu} 
\affiliation{Department of Physics \& Astronomy, University of Missouri, Columbia,
  MO 65211}

\date{\today}
\begin{abstract}
  The notion of tissue surface tension has provided a physical understanding of
  morphogenetic phenomena such as tissue spreading or cell sorting. The
  measurement of tissue surface tension so far relied on strong approximations on
  the geometric profile of a spherical droplet compressed between parallel plates.
  We solved the Laplace equation for this geometry and tested its solution on
  true liquids and embryonic tissue fragments as well as multicellular aggregates.
  The analytic solution provides the surface tension in terms of easily and
  accurately measurable geometric parameters. Experimental results show that the
  various tissues and multicellular aggregates studied here are incompressible
  and, similarly to true liquids, possess effective surface tensions that are
  independent of the magnitude of the compressive force and the volume of the
  droplet.
\end{abstract}

\maketitle

In the absence of external forces, an embryonic tissue fragment (typically less
than $1$~mm in size) rounds up similarly to a liquid drop, to minimize its surface
energy. Cells of two distinct tissues randomly intermixed within a single
multicellular aggregate sort into separate regions similarly to coalescing
immiscible liquids. 
To account for these observations Steinberg formulated the Differential Adhesion
Hypothesis (DAH), which states that embryonic tissues or, more generally, tissues
composed of motile cells with different homotypic adhesive strengths should behave
analogously to immiscible liquids \cite{steinberg70-395,Steinberg82}. DAH implies
that such tissues, similarly to ordinary liquids, possess measurable surface and
interfacial tensions generated by adhesive and cohesive interactions between the
component subunits (molecules in one case, cells in the other).
Predictions of DAH have been confirmed both in vitro \cite{foty96-1611} and in
vivo \cite{godt98-387,gonzalez-reyes98-3635,hayashi04-647}, and the surface
tensions of different embryonic tissues were measured and the values accounted for
the observed mutual sorting behavior \cite{foty94-2298,foty96-1611}.
Currently, the only available method to measure the surface or interfacial tension
$\sigma$ of submillimeter size droplets of tissue aggregates is by compression
plate tensiometry \cite{foty94-2298,forgacs98-2227}. The method, based on the Laplace
equation, so far relied on various approximations of the geometrical profile of an
equilibrated spherical droplet compressed between two parallel plates and yielded
$\sigma$ values that can at best only be considered relative and their
independence on droplet size and compressive force questionable.
In this Letter, by analytically solving the corresponding Laplace equation, we
determine the exact profile of a compressed droplet, which allows, for the first
time, to accurately and reliably determine the absolute value of tissue surface
tensions in terms of easily and accurately measurable geometric parameters.
Furthermore, we show that the method can readily be extended to the simultaneous
compression of several droplets of different sizes. 
Finally, we apply our new method to a large number of compression plate
measurements for calculating $\sigma$ of true liquids (for validation) and several
tissue and multicellular aggregates. 
Our results show that the studied systems are incompressible and the obtained
surface tensions are independent of the magnitude of the compressive force and the
volume of the droplets. Our results provide strong evidence for the concept of embryonic
tissue liquidity  and support its usefulness for the interpretation of early
morphogenetic processes.
More importantly, since surface tension is a measure of the liquid's cohesivity,
in the case of tissues it must be related to molecular parameters. Indeed, it was
shown on theoretical grounds that $\sigma \propto JN\tau$, where $J$, $N$ and
$\tau$ are respectively the bond energy between two homotypic cell adhesion
molecules (CAMs), the surface density of CAMs and the effective life time of the
adhesive bond \cite{forgacs98-2227}. The linear dependence of $\sigma$ on $N$ has
recently been confirmed experimentally \cite{foty05-255}. Thus our method of
determining the absolute value of $\sigma$ has important biological implications
as it quantitatively relates a macroscopic tissue property to biomolecular
entities.

\begin{figure}[t]
  \begin{center} 
\includegraphics[width=3.25in]{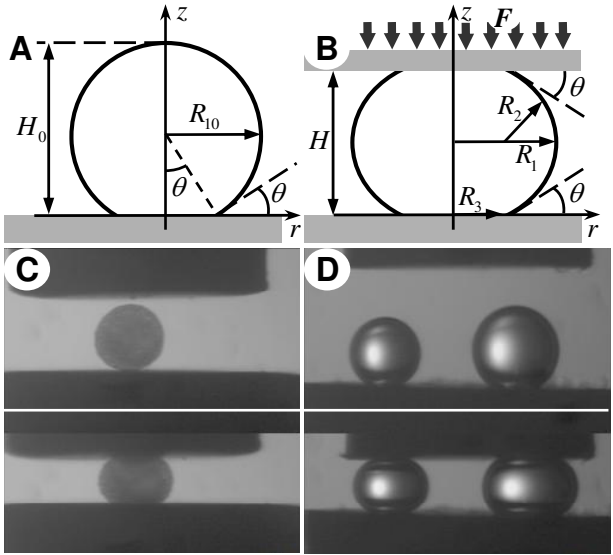}   
  \end{center}
  \caption{Diagram of an (A) uncompressed and (B) compressed liquid drop. (C)
    Snapshots of uncompressed (top) and compressed (bottom) cushion tissue (CT)
    droplet in culture medium. (D) Same for two different-size water drops in
    olive oil.}
\label{fig:fig1}
\end{figure}

The surface tension $\sigma$ of a small liquid droplet compressed between two
pressure plates can be determined from its geometric shape
(Fig.~\ref{fig:fig1}A-B).  Here we consider droplets with radius $R_0$ much
smaller than the corresponding capillary length $R_c\approx (\sigma/\rho
g)^{1/2}$, for which the effect of gravity can be neglected (e.g., for water, with
$\sigma =0.07\,\text{N/m}$ and density $\rho=10^3 \text{kg/m}^3$, $R_c \approx
2.7\,\text{mm})$. Thus, the shape of a submillimeter liquid drop placed on a
horizontal plate (Fig.~\ref{fig:fig1}A) is a spherical cap of radius $R_{10}$ and
height $H_{0}$.
With these parameters, simple geometric considerations provide: (i) the
(complementary) contact angle $\theta =\cos^{-1}\left(H_0/R_{10}-1\right)$, and,
assuming incompressibility, (ii) the radius of the suspended drop, $R_0=R_{10}[(2
- \cos{\theta})\cos^4(\theta/2)]$.
While $H_0$ and $R_{10}$ can be measured with high accuracy (e.g., $<1$\%), the
relative error $\Delta\theta/\theta \approx [(1+\cos\theta)/\theta \sin\theta]
(\Delta{H_0}/H_0 + \Delta{R_{10}}/R_{10})$ can still be very large (e.g., $\sim
30$\% for $\theta=20^{\circ}$ and $\gg 130$\% for $\theta<10^{\circ}$). 
Thus, for determining the surface tension it is desirable to reduce the adherence
between the drop and plates and avoid using quantities that explicitly contain the
contact angle.

The compressed drop (Fig.~\ref{fig:fig1}B) has rotational symmetry about the
$z$-axis and reflection symmetry with respect to the equatorial plane $z=H/2$. In
this plane the surface of the drop has two principal radii of curvatures $R_1$ and
$R_2$ (Fig.~\ref{fig:fig1}B).  $R_3$ is the radius of the droplet's circular area
of contact with either compression plates. The degree of compression depends on
the magnitude of the compression force $F$ applied to the upper (or lower) plate.
In terms of $R_1$ and $R_2$ the excess pressure inside the drop due to the surface
tension is given by the Laplace formula $\Delta{p}=\sigma (1/R_1+1/R_2)$.
Thus, at mechanical equilibrium, the balance of forces dictates the following
relations, respectively valid at the upper (or lower) plate and the equator
\begin{subequations}
\begin{equation}
  \label{eq:1a}
  \begin{split}
  F &= \Delta{p}\, \pi R_3^2 - 2 \pi R_3\sigma\sin\theta \\
    &= \pi\sigma \left[R_3^2(1/R_1+1/R_2)-2 R_3\sin\theta \right]\;,
  \end{split}
\end{equation}
\begin{equation}
  \label{eq:1b}
  F = \Delta{p}\, \pi R_1^2 - 2\pi R_1 \sigma = \pi\sigma R_1(R_1/R_2 -1) \;.
\end{equation}
\label{eq:1}
\end{subequations}
\noindent Although $H$, $R_1 $ and $F$ can be easily measured, the accurate
determination of $\sigma$ directly from either of these two equations also
requires the quantities $R_2$, $\theta$ and $R_3$ that can only be measured with
large errors (especially the latter two). 
This problem is often circumvented by assuming $\theta =0$ (no adhesion to the
plates) and/or making approximations on the lateral profile of the drop, e.g.,
$R_3=R_1-R_2$ (the profile is a semicircle) or $R_3=R_1-R_2+[R_2^2-(H/2)^2]^{1/2}$ (the
profile is a circular arc) \cite{foty96-1611,forgacs98-2227}.  Each of these
schemes fails to give consistent results in some range of the compressive force or
contact angle. For example approximating the lateral profile with a circular arc
implies that the contact angle depends on the magnitude of the compressive force,
a physically absurd conclusion.

Clearly, the precise determination of the radii $R_2$ and $R_3$ requires the
knowledge of the exact profile $z(r)$ of the compressed liquid droplet
(Fig.~\ref{fig:fig1}B). This can be obtained by the exact integration of the
Laplace equation
\begin{equation}
\label{eq:2}
\Delta p = \sigma \left[ 
  {\frac{{z}'}{r(1+{z}'^2)^{1/2}}+\frac{{z}''}{(1+{z}'^2)^{3/2}}} 
\right]=\text{const},
\end{equation}
subject to the boundary conditions (see Fig.~\ref{fig:fig1}B) 
\begin{equation}
\label{eq:3}
\begin{split}
z(R_3 )&=0,   \qquad z'(R_3 )=\tan \theta ,\\
z(R_1 )&=H/2, \quad z'(R_1 )=\infty \;.
\end{split}
\end{equation}
The terms in the square brackets in Eq.~\eqref{eq:2} represent the principal
curvatures of the drop's surface at a point determined by $z(r)$.

We are now in the position to determine $\sigma$ in terms of the easily and
accurately measurable quantities $H$, $R_1$ and $F$. Equations~\eqref{eq:1} imply
that both $R_2$ and $R_3$ can be expressed in terms of $R_1$
\begin{subequations}
\label{eq:R23ab}
  \begin{equation}
  \label{eq:R23}
    R_2 = R_1/(2\alpha -1)\;, \quad R_3=\beta R_1\;, 
  \end{equation}
where the dimensionless parameters $\alpha$ and $\beta$ are given by
\begin{equation}
  \label{eq:alpha}
  \alpha = \Delta{p}/(2\sigma/R_1)\;,
\end{equation}
  \begin{equation}
  \label{eq:beta}
    \beta \equiv \beta _\theta (\alpha )=\left( {2\alpha } \right)^{-1}\left[
      {\sin \theta +\sqrt {\sin ^2\theta +4\alpha (\alpha -1)} } \right]\;.
\end{equation}
\end{subequations}
Integrating Eq.~\eqref{eq:2} with the boundary conditions \eqref{eq:3} leads to an
implicit equation for $\alpha$
\begin{equation}
\label{eq:4}
\begin{split}
  \frac{H}{2R_1 }&=f_\theta (\alpha )\equiv \int\limits_\beta ^1 z'(x)dx\;,\\
  z'(x) &=\left[ {\left( {\frac{x}{\alpha x^2+1-\alpha }} \right)^2-1}
\right]^{-1/2}\;,
\end{split}
\end{equation}
and to the lateral profile of the compressed drop
\begin{equation}
  \label{eq:z-R}
  z(r)= R_1 \int_{\beta}^{r/R_1}z'(x) dx\;.
\end{equation}

\noindent Because the value of the integral in Eq.~\eqref{eq:4} depends weakly on
the lower integration limit $\beta$, one finds that $\alpha$, and therefore $R_2$,
are relatively insensitive to even large ($\sim 10^{\circ}$) sample-to-sample
fluctuations of $\theta$ (e.g., caused by local inhomogeneities or impurities). 
By contrast, since $\beta $ itself strongly depends on $\theta $, so does $R_3$.
This explains why its determination by fitting the lateral profile of the drop by
circular arcs is impractical.

Finally, $\sigma$ can be expressed from either Eqs.~\eqref{eq:1}, which in terms
of $\alpha \approx f_0^{-1}(H/2R_1)$ [see Eq.~\eqref{eq:4}] can be rewritten as
\begin{equation}
  \label{eq:5}
  \lambda_0\equiv F/2\pi R_1\sigma_0 = (\sigma/\sigma_0) (\alpha -1)\;, \qquad 
  \sigma_0 = 1~\text{mN/m} \;.
\end{equation}
This result provides a simple recipe to evaluate $\sigma$ from the measurement of
$H$, $R_1$ and $F$.  
The efficiency of the proposed method can be further enhanced
by simultaneously compressing several drops as shown in Fig.~\ref{fig:fig1}D. 
For such compressions the quantities $\lambda_0$ and $\alpha$ in Eq.~\eqref{eq:5}
need to be replaced respectively by $F/2\pi\sigma_0\bar{R}_1$ and
$\bar{R}_1^{-1}\sum_nR_{1n}\alpha_n$, with $\bar{R}_1=\sum_n R_{1n}$ and
$\alpha_n=f_0^{-1}(H/2R_{1n})$.

It should be emphasized that Eq.~\eqref{eq:4} is valid for incompressible and 
compressible liquids alike. In addition, for incompressible liquids volume 
conservation yields 
\begin{equation}
\label{eq:6}
({R_0 } \mathord{\left/ {\vphantom {{R_0 } {R_1 }}} \right. 
\kern-\nulldelimiterspace} {R_1 })^3=g_\theta (\alpha )\equiv 
\int_\beta^1 {x^2{z}'(x)dx} \; .
\end{equation}
For $\theta \le 20^{\circ}$, to a very good approximation, $g_{\theta}(\alpha)
\approx g_0(\alpha)$. Eliminating $\alpha$ between Eqs.~\eqref{eq:4} and
\eqref{eq:6} leads to
\begin{equation}
  \label{eq:7}
  H/2R_1 = U_{\theta}(H/2R_0) \approx U_0(H/2R_0)\;, 
\end{equation}
where $U_0$ is a universal function determined by the functions $f_0 $ and $g_0$
\cite{U0}.  Thus, when the adhesion between the drop and plates is weak (i.e.,
$\theta\le 20^{\circ}$) there is a universal relationship between $H/2R_1$ and
$H/2R_0$, valid for any incompressible liquid drop regardless of its type or size.

\begin{figure}
  \centering
  \includegraphics[width=3.25in]{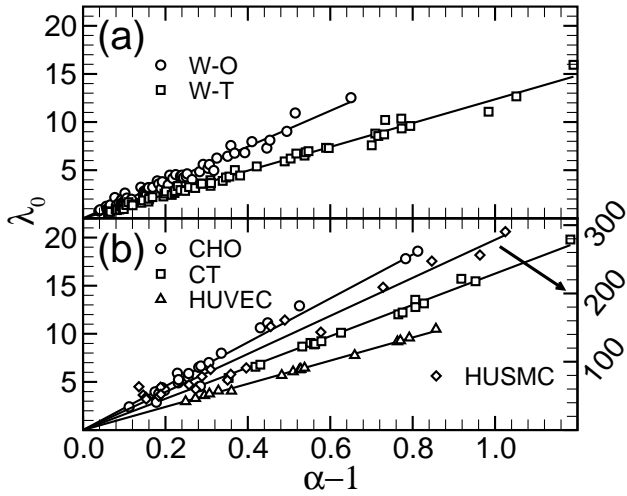}
  \caption{Implementation of Eq.~\eqref{eq:5} to evaluate the experimental data
    obtained in the compression measurements. The surface and interfacial tensions
    are obtained from the slopes of the linear fit to the data points including
    the origin: (a) pure liquid (W-O, W-T), and (b) tissues and multicellular
    aggregates (CT, CHO, HUVEC, HUSMC).}
  \label{fig:2}
\end{figure}

\begin{table}
  \centering
  \caption{Surface and interfacial tensions of the studied systems obtained from
    the data shown in Fig.~\ref{fig:2}. The absolute errors
    $\Delta\sigma$ are standard deviations. The last two columns contain
    respectively the percentage relative errors and the number of data points.}
  \begin{ruledtabular}
  \begin{tabular}{ccccc}
    system & $\sigma$ & $\Delta\sigma$ & $\Delta\sigma/\sigma$ & data  \\
           & [mN/m] & [mN/m] & [\%] & points \\
    \hline
    W-O   & 18.6 & 2.4 & 12.9 & 56 \\
    W-T   & 12.4 & 1.5 & 12.1 & 60 \\
    CHO   & 22.8 & 3.0 & 13.2 & 22 \\
    CT    & 16.3 & 0.2 & 1.2  & 17 \\
    HUVEC & 12.0 & 0.2 & 1.7  & 16 \\
    HUSMC & 279  & 57  & 20.4 & 21 \\
  \end{tabular}
  \end{ruledtabular}
  \label{tab:1}
\end{table}

We have tested the above theory to determine the interfacial tension of true
immiscible liquids (water in olive oil (W-O) and turpentine (W-T)), where results
obtained by other methods are available. Subsequently, we applied the analytical
results to determine the absolute values of tissue surface tensions. Measurements
were performed using a specifically designed compression tensiometer (for details
see \cite{hegedus06-2708}). In the case of ordinary liquids, one or two spherical
water drops of various size (ranging from $0.5$ to $1$~mm in diameter) were
compressed simultaneously (Fig.~\ref{fig:fig1}D). In each experiment, drops were
exposed to at least 3 successive compressions (up to 9) of increasing force and
their profile was recorded at shape equilibrium.

\begin{figure}
  \centering
  \includegraphics[width=3.25in]{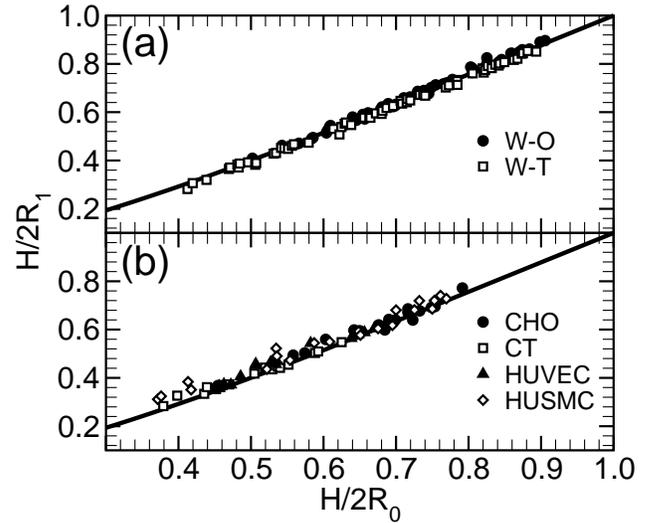}
  \caption{Implementation of the universal function given in Eq.~\eqref{eq:7}
    (solid curve) to assess the incompressibility of the (a) liquid and (b) tissue
    and multicellular droplets used in the compression measurements.}
  \label{fig:3}
\end{figure}

For biological measurements we either used intact tissue (fragments of excised
embryonic chicken cardiac cushions (CT) that round into spheres in about 12 hours)
or model tissues (spherical aggregates composed of various cell types, such as
Human Umbilical Smooth Muscle Cells (HUSMC), Human Umbilical Vein Endothelial
Cells (HUVEC) and Chinese Hamster Ovary cells (CHO)).  Multicellular aggregates
were prepared as previously described \cite{hegedus06-2708}. Tissue droplets were
compressed in culture medium at 37$^{\circ}$C (Fig.~\ref{fig:fig1}C). One to six
droplets of each type (ranging from $250$ to $600~\mu$m in diameter) were
compressed simultaneously. In order to avoid irreversible damage to the cells, no
more than 2 compressions were performed on each droplet.

According to the solution of the Laplace equation \eqref{eq:5}, for liquids, the
pairs of data points $\{(\alpha-1),\lambda_0\}$ should lie on a straight line
passing through the origin, with $\sigma$ given by the slope of the line. 
The data points are shown in Fig.~\ref{fig:2} and the obtained results are
summarized in Table~\ref{tab:1}. Besides the values of $\sigma$, the table also
contains the standard deviation ($\Delta\sigma$) of the measurements and the
corresponding percentage relative errors ($\Delta\sigma/\sigma$). The last column
of the table lists for each system the total number of data points used and shown
in Fig.~\ref{fig:2}.
The obtained interfacial tensions of W-O ($18.6\pm 2.4$~mN/m) and W-T ($12.4\pm
1.5$ mN/m) compare rather well with published data \cite{nouy25-625}, i.e., $15.4$
and $13.7$~mN/m, respectively.
It should be noted that the corresponding relatively large errors ($~12\%$) in
Table~\ref{tab:1} reflect the sample sensitivity of $\sigma$ and not a deficiency
of the method to determine it.  Indeed, the large number of W-O and W-T
compression measurements (see Table~\ref{tab:1}) were done in the course of four
different days, each time using different samples and compression plates.  Also,
most of these experiments were carried out using single droplets, which might have
experienced somewhat different environmental conditions (sample-to-sample
fluctuations).  The actual errors $\Delta\sigma$ corresponding to drops from the
same batch were much smaller ($\sim 1\%$).
Another possible source of error is related to the extent of compressions. In
general, results from weaker compressions have larger errors. Indeed, the relative
error of the surface tension $\Delta\sigma/\sigma = \Delta{F}/F + \Delta{R_1}/R_1
+ \Delta\alpha/(\alpha-1)$ [obtained from Eq.~\eqref{eq:5}] may become very large
for small compressions when $F\rightarrow 0$ and $\alpha\rightarrow 1$.

By a similar procedure, the surface tensions of tissue droplets (i.e., interfacial
tension in tissue culture medium) were also determined with the results shown in
Fig.~\ref{fig:2}b and Table~\ref{tab:1}. The fact that the data points in
Fig.~\ref{fig:2}b corresponding to each type of cell aggregates fall on well
defined lines that go through the origin clearly demonstrates that embryonic
tissues and cell aggregates have well defined surface tensions that are
independent of their size or extent of compression.
The small $\Delta\sigma$ in case of CT (HUVEC) aggregates is due to the fact that
the measurements were carried out using a single batch (of simultaneously
prepared) aggregates. These aggregates were exposed to multiple multi-aggregate
compressions: five or six aggregates compressed twice.
In the case of CHO aggregates originating from a single batch, where a bigger
number of compressions were performed on both single and multiple (two to five)
aggregates, the relative error $\Delta\sigma/\sigma$ was similar to that of W-O
and W-T.
In the case of HUSMC four different batches of aggregates were used and only 
single aggregate compressions were performed. This may explain the 
relatively large error ($\sim 20\%$).
Thus one may conclude that multi-aggregate compressions with droplets originating
from a single batch is the most efficient and accurate way of experimentally
determining tissue surface tension.  This is particularly important for highly
viscous aggregates of living cells whose properties may change during their
relatively long relaxation time in a compression measurement.

Furthermore, the fact that the data points $\{H/2R_0,H/2R_1\}$ for all our
measurements fall on the universal curve predicted by Eq.~\eqref{eq:7}
(Fig.~\ref{fig:3}) demonstrates that the embryonic tissues and multicellular
aggregates studied here, similarly to water, are incompressible.

The results in Figs.~\ref{fig:2} and \ref{fig:3} provide a wealth of information.
They convincingly demonstrate that embryonic tissues or multicellular aggregates
composed of embryonic cells (model tissues) indeed manifest liquid-like
properties.  Thus, such tissues are incompressible and can quantitatively be
characterized in terms of liquid-like surface tensions. The measured values of
these parameters are independent of the size of the drops used or the magnitude of
the force exerted on them. The employed method therefore provides the absolute
values of tissue surface tensions. 
As discussed earlier, these results can be used to assess biomolecular quantities.

The analytic solution of the Laplace equation presented here, combined with
compression plate tensiometry, provides a novel, reliable and accurate way to
determine liquid surface tension (and to our knowledge the only method applicable
to tissues). Our approach may be particularly useful in the case when the
interfacial tension of expensive liquids needs to be determined, where only small
quantities of the materials are available.

This work was supported by the National Science Foundation [FIBR-0526854].


\begin{thebibliography}{13}
\expandafter\ifx\csname natexlab\endcsname\relax\def\natexlab#1{#1}\fi
\expandafter\ifx\csname bibnamefont\endcsname\relax
  \def\bibnamefont#1{#1}\fi
\expandafter\ifx\csname bibfnamefont\endcsname\relax
  \def\bibfnamefont#1{#1}\fi
\expandafter\ifx\csname citenamefont\endcsname\relax
  \def\citenamefont#1{#1}\fi
\expandafter\ifx\csname url\endcsname\relax
  \def\url#1{\texttt{#1}}\fi
\expandafter\ifx\csname urlprefix\endcsname\relax\def\urlprefix{URL }\fi
\providecommand{\bibinfo}[2]{#2}
\providecommand{\eprint}[2][]{\url{#2}}

\bibitem[{\citenamefont{Steinberg}(1970)}]{steinberg70-395}
\bibinfo{author}{\bibfnamefont{M.~S.} \bibnamefont{Steinberg}},
  \bibinfo{journal}{J Exp Zool} \textbf{\bibinfo{volume}{173}},
  \bibinfo{pages}{395} (\bibinfo{year}{1970}).

\bibitem[{\citenamefont{Steinberg and Poole}(1982)}]{Steinberg82}
\bibinfo{author}{\bibfnamefont{M.}~\bibnamefont{Steinberg}} \bibnamefont{and}
  \bibinfo{author}{\bibfnamefont{T.}~\bibnamefont{Poole}}, in
  \emph{\bibinfo{booktitle}{Cell Behaviour}}, edited by
  \bibinfo{editor}{\bibfnamefont{R.}~\bibnamefont{Bellairs}},
  \bibinfo{editor}{\bibfnamefont{A.}~\bibnamefont{Curtis}}, \bibnamefont{and}
  \bibinfo{editor}{\bibfnamefont{G.}~\bibnamefont{Dunn}}
  (\bibinfo{publisher}{Cambridge Unversity Press},
  \bibinfo{address}{Cambridge}, \bibinfo{year}{1982}), pp.
  \bibinfo{pages}{583--607}.

\bibitem[{\citenamefont{Foty et~al.}(1996)\citenamefont{Foty, Pfleger, Forgacs,
  and Steinberg}}]{foty96-1611}
\bibinfo{author}{\bibfnamefont{R.~A.} \bibnamefont{Foty}},
  \bibinfo{author}{\bibfnamefont{C.~M.} \bibnamefont{Pfleger}},
  \bibinfo{author}{\bibfnamefont{G.}~\bibnamefont{Forgacs}}, \bibnamefont{and}
  \bibinfo{author}{\bibfnamefont{M.~S.} \bibnamefont{Steinberg}},
  \bibinfo{journal}{Development} \textbf{\bibinfo{volume}{122}},
  \bibinfo{pages}{1611} (\bibinfo{year}{1996}).

\bibitem[{\citenamefont{Godt and Tepass}(1998)}]{godt98-387}
\bibinfo{author}{\bibfnamefont{D.}~\bibnamefont{Godt}} \bibnamefont{and}
  \bibinfo{author}{\bibfnamefont{U.}~\bibnamefont{Tepass}},
  \bibinfo{journal}{Nature} \textbf{\bibinfo{volume}{395}},
  \bibinfo{pages}{387} (\bibinfo{year}{1998}).

\bibitem[{\citenamefont{Gonzalez-Reyes and
  St~Johnston}(1998)}]{gonzalez-reyes98-3635}
\bibinfo{author}{\bibfnamefont{A.}~\bibnamefont{Gonzalez-Reyes}}
  \bibnamefont{and}
  \bibinfo{author}{\bibfnamefont{D.}~\bibnamefont{St~Johnston}},
  \bibinfo{journal}{Development} \textbf{\bibinfo{volume}{125}},
  \bibinfo{pages}{3635} (\bibinfo{year}{1998}).

\bibitem[{\citenamefont{Hayashi and Carthew}(2004)}]{hayashi04-647}
\bibinfo{author}{\bibfnamefont{T.}~\bibnamefont{Hayashi}} \bibnamefont{and}
  \bibinfo{author}{\bibfnamefont{R.~W.} \bibnamefont{Carthew}},
  \bibinfo{journal}{Nature} \textbf{\bibinfo{volume}{431}},
  \bibinfo{pages}{647} (\bibinfo{year}{2004}).

\bibitem[{\citenamefont{Foty et~al.}(1994)\citenamefont{Foty, Forgacs, Pfleger,
  and Steinberg}}]{foty94-2298}
\bibinfo{author}{\bibfnamefont{R.~A.} \bibnamefont{Foty}},
  \bibinfo{author}{\bibfnamefont{G.}~\bibnamefont{Forgacs}},
  \bibinfo{author}{\bibfnamefont{C.~M.} \bibnamefont{Pfleger}},
  \bibnamefont{and} \bibinfo{author}{\bibfnamefont{M.~S.}
  \bibnamefont{Steinberg}}, \bibinfo{journal}{Phys Rev Lett}
  \textbf{\bibinfo{volume}{72}}, \bibinfo{pages}{2298} (\bibinfo{year}{1994}).

\bibitem[{\citenamefont{Forgacs et~al.}(1998)\citenamefont{Forgacs, Foty,
  Shafrir, and Steinberg}}]{forgacs98-2227}
\bibinfo{author}{\bibfnamefont{G.}~\bibnamefont{Forgacs}},
  \bibinfo{author}{\bibfnamefont{R.~A.} \bibnamefont{Foty}},
  \bibinfo{author}{\bibfnamefont{Y.}~\bibnamefont{Shafrir}}, \bibnamefont{and}
  \bibinfo{author}{\bibfnamefont{M.~S.} \bibnamefont{Steinberg}},
  \bibinfo{journal}{Biophys J} \textbf{\bibinfo{volume}{74}},
  \bibinfo{pages}{2227} (\bibinfo{year}{1998}).

\bibitem[{\citenamefont{Foty and Steinberg}(2005)}]{foty05-255}
\bibinfo{author}{\bibfnamefont{R.~A.} \bibnamefont{Foty}} \bibnamefont{and}
  \bibinfo{author}{\bibfnamefont{M.~S.} \bibnamefont{Steinberg}},
  \bibinfo{journal}{Dev Biol} \textbf{\bibinfo{volume}{278}},
  \bibinfo{pages}{255} (\bibinfo{year}{2005}).

\bibitem[{U0()}]{U0}
\bibinfo{note}{The universal function $U_0$ has the following parametric form:
  $U_0(x)=f_0(\alpha)$ and $x=f_0(\alpha)[g_0(\alpha)]^{1/3}$.}

\bibitem[{\citenamefont{Hegedus et~al.}(2006)\citenamefont{Hegedus, Marga,
  Jakab, Sharpe-Timms, and Forgacs}}]{hegedus06-2708}
\bibinfo{author}{\bibfnamefont{B.}~\bibnamefont{Hegedus}},
  \bibinfo{author}{\bibfnamefont{F.}~\bibnamefont{Marga}},
  \bibinfo{author}{\bibfnamefont{K.}~\bibnamefont{Jakab}},
  \bibinfo{author}{\bibfnamefont{K.~L.} \bibnamefont{Sharpe-Timms}},
  \bibnamefont{and} \bibinfo{author}{\bibfnamefont{G.}~\bibnamefont{Forgacs}},
  \bibinfo{journal}{Biophys J} \textbf{\bibinfo{volume}{91}},
  \bibinfo{pages}{2708} (\bibinfo{year}{2006}).

\bibitem[{\citenamefont{{du Nouy}}(1925)}]{nouy25-625}
\bibinfo{author}{\bibfnamefont{P.~L.} \bibnamefont{{du Nouy}}},
  \bibinfo{journal}{J. Gen. Physiol.} \textbf{\bibinfo{volume}{7}},
  \bibinfo{pages}{625} (\bibinfo{year}{1925}).

\end{thebibliography}

\end{document}